# Time-Encoded Raman: Fiber-based, hyperspectral, broadband stimulated Raman microscopy

Sebastian Karpf[1], Matthias Eibl[1], Wolfgang Wieser[1], Thomas Klein[1], Robert Huber[1,2,*]

[1]Lehrstuhl für BioMolekulare Optik, Fakultät für Physik, Ludwig-Maximilians-Universität München, Oettingenstr. 67, 80538 Munich, Germany
[2]Institut für Biomedizinische Optik, Universität zu Lübeck, Peter-Monnik-Weg 4, 23562 Lübeck
*Correspondence to: robert.huber@bmo.uni-luebeck.de

**Raman sensing and Raman microscopy are amongst the most specific optical technologies to identify the chemical compounds of unknown samples, and to enable label-free biomedical imaging with molecular contrast. However, the high cost and complexity, low speed, and incomplete spectral information provided by current technology are major challenges preventing more widespread application of Raman systems. To overcome these limitations, we developed a new method for stimulated Raman spectroscopy and Raman imaging using continuous wave (CW), rapidly wavelength swept lasers. Our all-fiber, time-encoded Raman (TICO-Raman) setup uses a Fourier Domain Mode Locked (FDML) laser source to achieve a unique combination of high speed, broad spectral coverage (750 cm$^{-1}$ - 3150 cm$^{-1}$) and high resolution (0.5 cm$^{-1}$). The Raman information is directly encoded and acquired in time. We demonstrate quantitative chemical analysis of a solvent mixture and hyperspectral Raman microscopy with molecular contrast of plant cells.**

Raman spectroscopy is an optical technology directly sensitive to molecular vibrations. Therefore, it can provide a wealth of information about the chemical composition of a sample. In biomedical applications it enables label free molecular imaging *in vivo*. Being an optical technique, it can combine chemical contrast with high spatial resolution on a micron scale. Raman microscopy has great potential for diverse applications such as tumor detection *(1)*, drug delivery studies *(2, 3)*, and biofuel process monitoring *(4, 5)*.
Traditional linear Raman scattering suffers from inherently low signal, limiting its application mainly to non-imaging spectroscopy. The fundamental problem of the low signal can be overcome by non-linear techniques, which can enhance the Raman intensity by many orders of magnitude.
The most popular non-linear Raman techniques are CARS (coherent anti-stokes Raman scattering) and SRS (stimulated Raman scattering). With CARS, high signal-to-noise ratios have been achieved *(6)* and hyperspectral microscopy using a pulsed, stepwise wavelength-tuned laser has been demonstrated *(7)*. SRS systems using only one spectral position have pushed the speed up to video-rate *(8, 9)*. To achieve hyperspectral imaging, multi-color setups were developed *(2, 9-12)*. Recent efforts concentrate on fiber delivery *(13-16)* for future endoscopic applications *(17)*, and cost effective continuous wave (CW) systems *(18-20)*. However, the demand for a flexible, broadband, high resolution Raman imaging system based on fiber light sources is still unmet *(21)*. We present a new SRS technique based on fastest wavelength-swept CW probe lasers. With the well-defined time characteristic of the wavelength sweep of this laser, the spectral Raman information is encoded and acquired in time. We call this approach time-encoded Raman (TICO-Raman). This new technique can meet all the demands mentioned above; it is inherently low cost, readily compatible with endoscopes, and has full spectroscopic capabilities.

The TICO-Raman concept is sketched in Figure 1 (A-C). Two laser light sources, a Raman pump and a Raman probe laser, are focused onto the sample. A photodiode detects the probe power behind the sample (Fig. 1C). The pump laser operates at a fixed wavelength, while the probe light is repeatedly swept in wavelength over time (Fig. 1A). Therefore, the photon energy difference between the two lasers is changed periodically. Every time the difference matches a Raman transition of a sample molecule (Fig. 1B), pump photons will be converted to probe photons by SRS (Fig. 1C). Hence, the intensity of the probe laser at that specific spectral position gets increased by stimulated Raman gain (SRG). During a sweep of the probe laser, the probe signal will increase every time the wavelength difference coincides with a Raman transition (Fig. 1A-C). With the known time to wavelength encoding of the probe laser sweep, the SRG signals can be mapped to a Raman spectrum.

In our actual implementation (Fig. 1D) we use a slightly different setup in order to combine the great potential of CW-SRS *(18-20)* with the advantage of a pulsed pump laser with high peak intensity. Since the SRG scales linearly with the instantaneous pump power, we modulate the pump laser to low duty cycles in order to achieve high peak power levels at low average power. In order to generate Raman spectra, the timing of the pump pulses is successively increased with respect to the sweeps of the probe laser and the induced SRG signals are mapped to a spectrum (Fig. 1D).

As part of this effort we developed a pump laser that can generate three different wavelengths, which can be switched electronically. The pump laser is used in conjunction with two wavelength swept Fourier Domain Mode Locked (FDML) probe lasers, each FDML laser having different emission spectra. This setup can potentially yield gapless coverage of Raman transitions from 250 $cm^{-1}$ up to 3150 $cm^{-1}$.

The homebuilt fiber based pump laser (Fig. 2B) combines the concept of a fiber based master oscillator power amplifier (MOPA) *(22)* with a wavelength shifter in the delivery glass fiber *(23, 24)*. This allows to electronically switch the pump light wavelength from 1064 nm to 1122 nm or 1185 nm. Peak power, emission wavelength, duty cycle, repetition rate, pulse duration, and pulse pattern can be dynamically programmed. This pump source is fully compatible with fiber endoscopes, as self phase modulation (SPM) - the main nonlinear effect causing pulse break up in fibers - is reduced by about three orders of magnitude by the use of nanosecond pulses instead of commonly used picoseconds pulses.

The fiber based FDML probe laser (Fig. 2A) is a periodically wavelength swept ring laser *(25)*. It uses a km long fiber to optically store the wavelength sweep, making it a very low noise, repetitive wavelength swept CW light source.

FDML lasers achieve 150 nm wavelength sweep spans in the extended near infrared (exNIR) at high sweep repetition rates up to several MHz *(26)*. These sources are robust and are used for *in vivo* biomedical imaging in optical coherence tomography (OCT) *(26)*, including for applications in high-speed 3D endoscopic OCT *(27)*. Like the pump laser, FDML probe lasers are highly flexible. Wavelength sweep range and center wavelength can be freely reconfigured. The pump laser is electronically synchronized to the swept probe laser (Fig. 2D).

The intensity changes of the probe laser light due to SRG are small compared to the total intensity of the probe laser. Therefore, we apply a balanced detection scheme (Fig. 2C) to remove the probe light offset and, furthermore, eliminate laser noise. After this analog balancing step, a second, digital balancing scheme additionally removes artifacts, such as signals arising from acoustic waves, thermal lensing, and partial interference *(28)*. Together with the inherently low noise FDML laser we achieve fundamental shot noise limited detection sensitivity.

We investigated the advantages of the TICO-Raman system for two main applications: Raman spectroscopy and Raman microscopy.

First, the spectroscopy performance is demonstrated. The spectrum of a mixture of benzene, toluene and cyclohexane (Figure 3A) has broadband coverage from 750 $cm^{-1}$ to 3150 $cm^{-1}$, a high resolution of better than 3 $cm^{-1}$ and 1565 spectral points. Four spectral blocks (coverage Fig. 3A top) were averaged 1000 times and merged. The intensities were normalized by pump and probe power levels. The TICO-Raman spectra are in very good agreement with spontaneous Raman spectra (Fig. S7) *(29)*.

Another unique feature of TICO-Raman is the possibility of dynamically zooming into a spectral region of interest by reducing the span of the probe laser. The effect of increased spectral resolution is shown in Fig. 3A, where the resolution improved from 3 $cm^{-1}$ down to 0.5 $cm^{-1}$. The two narrowband, neighboring Raman peaks of toluene (1005 $cm^{-1}$) and benzene (992 $cm^{-1}$) are better resolved after zooming in. It should be noted that the resolution in TICO-Raman is not limited by a spectrometer, but by the linewidths of the lasers. Due to the pulse durations of nanoseconds, resolutions of GHz (~ 0.007 $cm^{-1}$) or better should be obtainable in the future *(20)*.

Another advantage of TICO-Raman, since it is an SRS technique, is the signal linearity in sample concentration. The left box in Figure 3B shows three individual TICO-Raman spectra of cyclohexane, benzene, and toluene. Figure 3B (right) shows a spectrum of the chemical mixture versus a spectrum calculated as weighted sum of the individual spectra. Both spectra agree very well, showing that TICO-Raman enables not only qualitative identification of the constituents of the examined sample but also their quantitative ratio.

As a second application of the TICO-Raman system, we performed hyperspectral SRS microscopy on plant cells by translating the sample. Figure 4 shows a slice of *geranium phaeum* stem immersed in olive oil with molecular contrast. We chose this sample as a model because the lignin distribution in plants is highly relevant for research on biomass-to-biofuel conversion, since it is a key factor in the recalcitrance process (3, 4). Olive oil was chosen as a representative of the group of lipid molecules.

At each sample location a 64-point spectrum from 1575 $cm^{-1}$ to 1665 $cm^{-1}$ was acquired. The power on the sample was 480 mW for the pump and 3.3 mW for the probe laser.

Depending on the application, the number of spectral points can be reconfigured dynamically for faster imaging speed.

Usually, non-linear imaging is performed in the 3000 cm$^{-1}$ region because of high signal levels, but the more specific region below 2000 cm$^{-1}$ would be preferred *(4)*.

We demonstrate the molecular distinction of lignin and olive oil around ~1600 cm$^{-1}$ in the fingerprint region. The narrowband vibrations are only 60 cm$^{-1}$ apart but clearly separable (Fig. 4B). Figure 4A shows the hyperspectral TICO-Raman image with lignin (red) and olive oil (green). At each pixel a spectrum was averaged 100 times, resulting in a pixel dwell time of 16 ms. The acquisition time for one averaged spectral point was 250 μs (2.5 μs non-averaged).

In Figure 4 the hyperspectral TICO-Raman contrast allows for molecular identification with high spatial resolution and high contrast. Figure 4B shows the spectra at pixels P1 and P2. The great advantage of numerous spectral points is evident, as lignin and olive oil can easily be identified and spectral integration over the Raman bands in post processing allows for optimal molecular contrast. The spectral integration regions are marked in green and red. The spectra were Savitzky-Golay filtered and offset corrected by a polynomial fit. The high specificity of TICO-Raman is shown in Figure 4C-D where each species is displayed individually.

The images in Figures 4A-D use the SRG signal of the probe beam. Additionally, the transmission of the pump beam can generate a confocal microscopy image. Mapping the molecular color code onto the high definition and high resolution morphology from the confocal image then provides the maximum amount of information in a single image (Fig. 4E). This multi-modal image allows to simultaneously identify the sample morphology with very high detail and the molecular composition with functional specificity.

Besides the advantages presented here, the TICO-Raman microscopy system additionally offers a straight forward extension to future multi-modal endoscopic imaging. Both lasers, TICO-Raman pump and TICO-Raman probe, can also be used for imaging modalities other than Raman. We have already demonstrated that FDML lasers can also be used for record high-speed endoscopic intravascular OCT imaging *(27)*. Additionally, the TICO-Raman pump laser has sufficient power levels for two photon microscopy (TPM) and second harmonic imaging (SHG). For *in vivo* OCT and TPM applications, the long wavelengths of both sources are very attractive for deep tissue imaging. Together with the robustness of the setup, TICO-Raman is the ideal candidate for molecularly sensitive endoscopy in a real clinical setting.

**Acknowledgments:** The authors acknowledge support from W. Zinth at the Ludwig-Maximilians-University Munich and the Munich Center for Advanced Photonics and D.C. Adler for fruitful discussion. This research was sponsored by the European Union project FDML-Raman (ERC, contract no. 259158).

**Figures:**

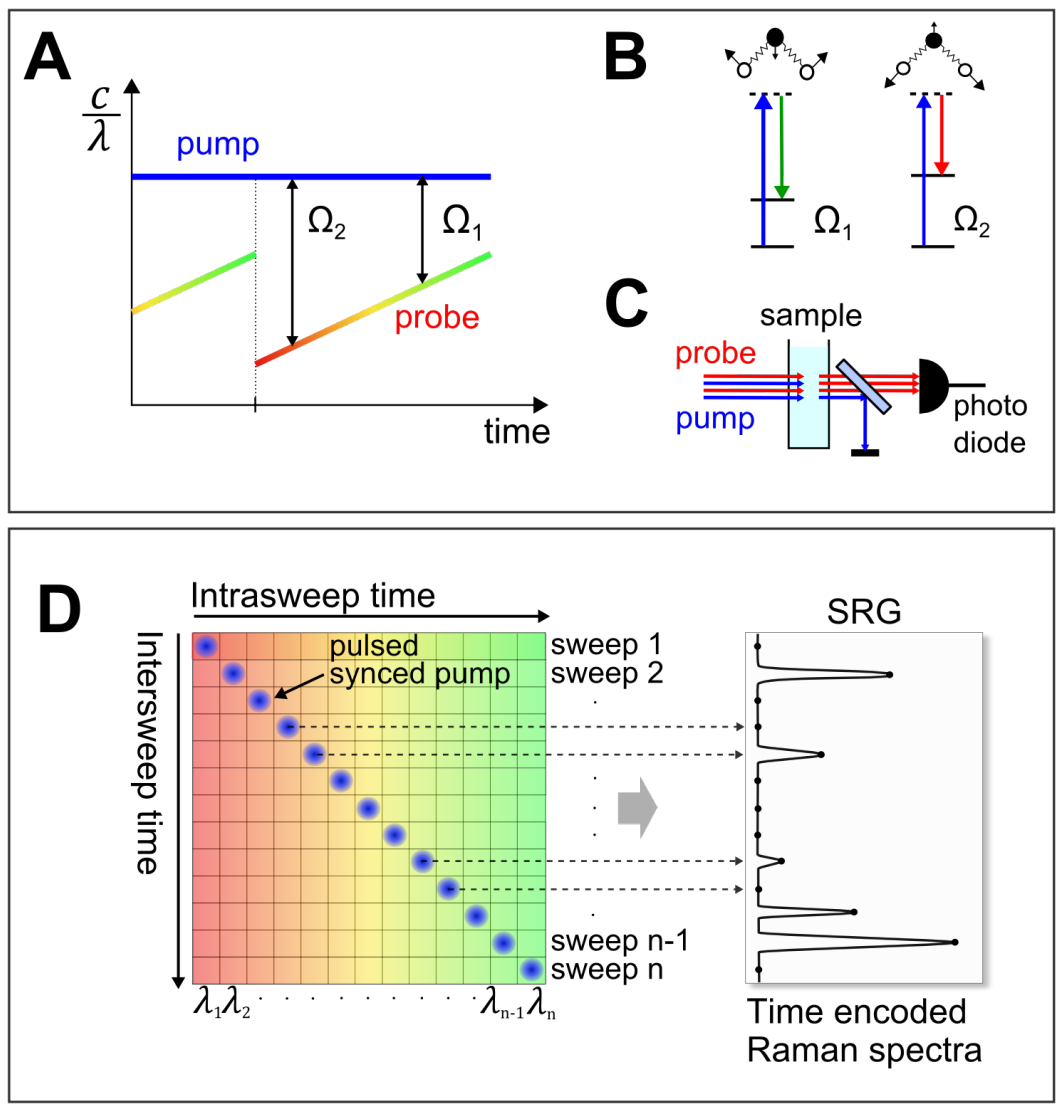

**Figure 1 Concept and implementation of TICO-Raman.** *The concept*: (**A**) A fixed wavelength pump and a swept wavelength probe laser scan molecular vibrations by changing the difference in photon energy. The probe laser experiences SRG when the energy difference matches Raman transitions ($\Omega_1$, $\Omega_2$). (**B**) Upon Raman scattering, pump photons are converted to probe photons and the loss in photon energy is transferred to molecular vibrations or rotations. (**C**) The SRG is measured as a signal change of the probe laser with a photodetector. *Actual implementation*: (**D**) The CW pump laser is modulated to nanosecond pulses which are synchronized to the CW probe laser. The probe laser sweeps the wavelength over time (intrasweep time, rainbow colored). TICO-Raman spectra are generated by positioning the pump pulses in time at probe wavelengths $\lambda_1,…,\lambda_n$ over consecutive sweeps 1,…,n. By direct time resolved sampling of the probe laser intensity, n SRG measurement points are acquired. With the known time-to-wavenumber characteristic, the n values are mapped to a Raman spectrum.

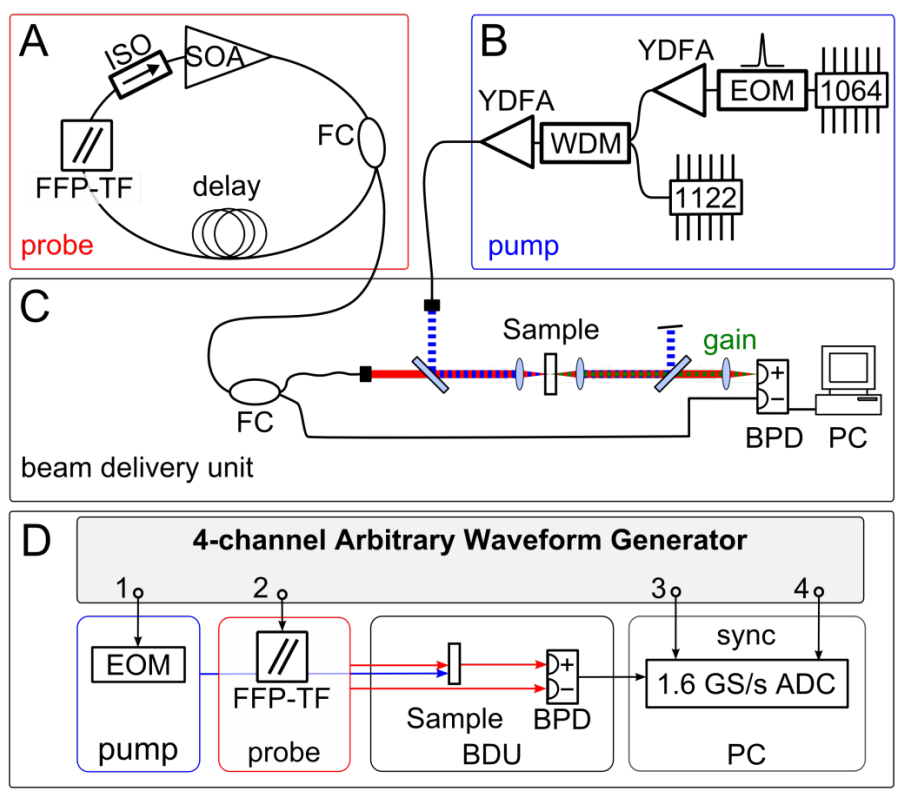

**Figure 2 The setup of the TICO-Raman system.** (**A**) The fiber based, wavelength swept FDML probe laser. SOA: semiconductor optical amplifier, ISO: optical isolator, FC: fiber coupler, FFP-TF: fiber Fabry-Pérot tunable filter. (**B**) The homebuilt fiber based pump laser is digitally synchronized to the FDML. EOM: electro-optic modulator, WDM: wavelength division multiplexer, YDFA: ytterbium doped fiber amplifier. (**C**) The lasers are combined in the beam delivery unit and focused onto the sample. The SRG signal is detected after subtraction of the offset by a differential balanced photodetector (BPD). (**D**) Digital synchronization is employed by an inter-channel locked arbitrary waveform generator driving the whole TICO-system. The SRG signals are directly sampled at 1.6 GS/s with a fast analog-to-digital converter card.

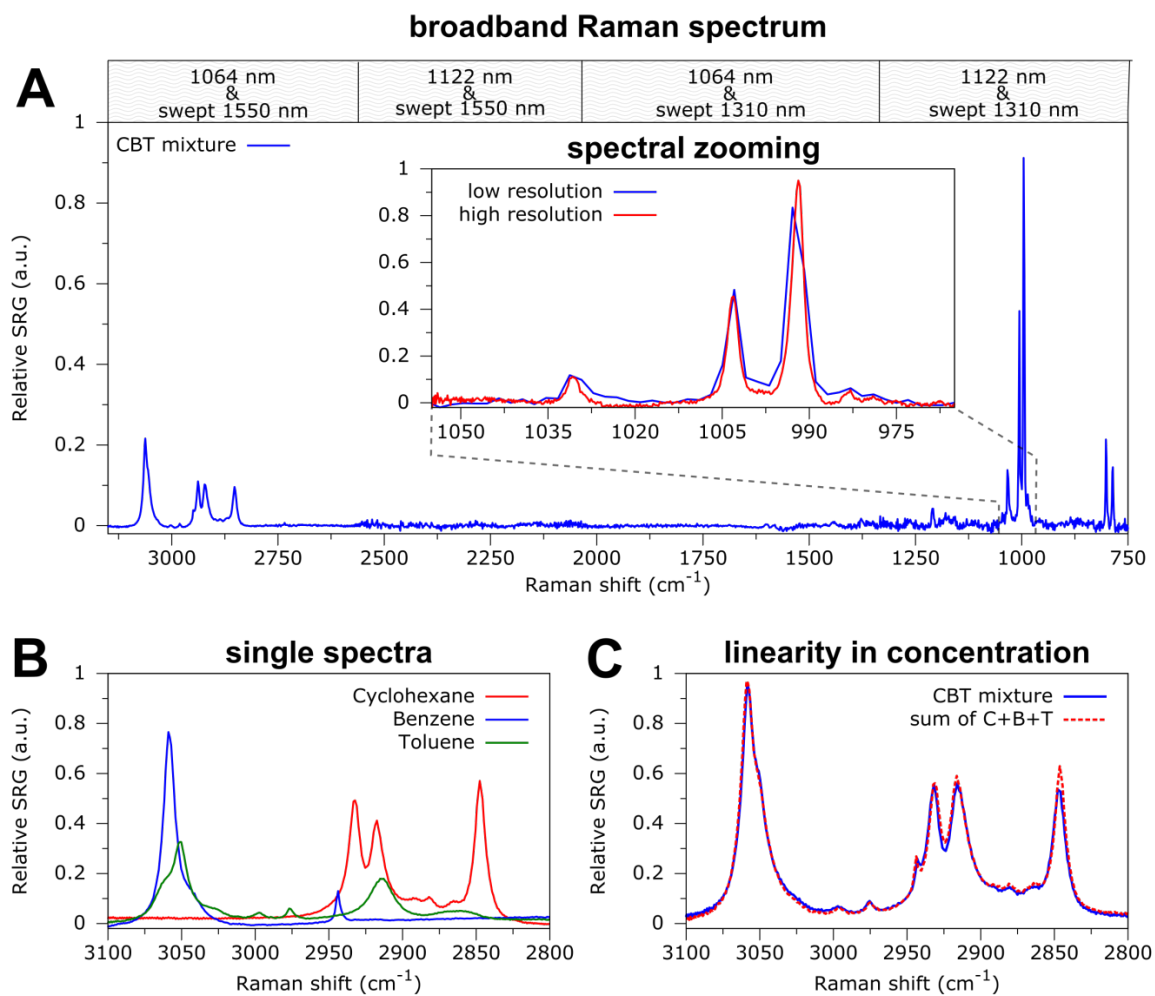

**Figure 3 Quantitative chemical sensing: mixture of cyclohexane, benzene and toluene (CBT, 1:1:1 ratio).** (**A**) Broadband TICO-Raman survey spectrum from 750 to 3150 cm$^{-1}$ with a resolution of 3 cm$^{-1}$. The spectrum has four sections acquired with two pump wavelengths and two FDML lasers. The inset shows the scope of dynamical spectral zooming. The sweep span of the swept laser is reduced, improving the spectral resolution to 0.5 cm$^{-1}$. The good linearity of the TICO-Raman signal enables quantitative determination of different compounds. (**B**) Spectra of the individual liquids. (**C**) The spectrum of the chemical mixture (blue) and the mathematical sum of the single spectra (red, dotted) match very well.

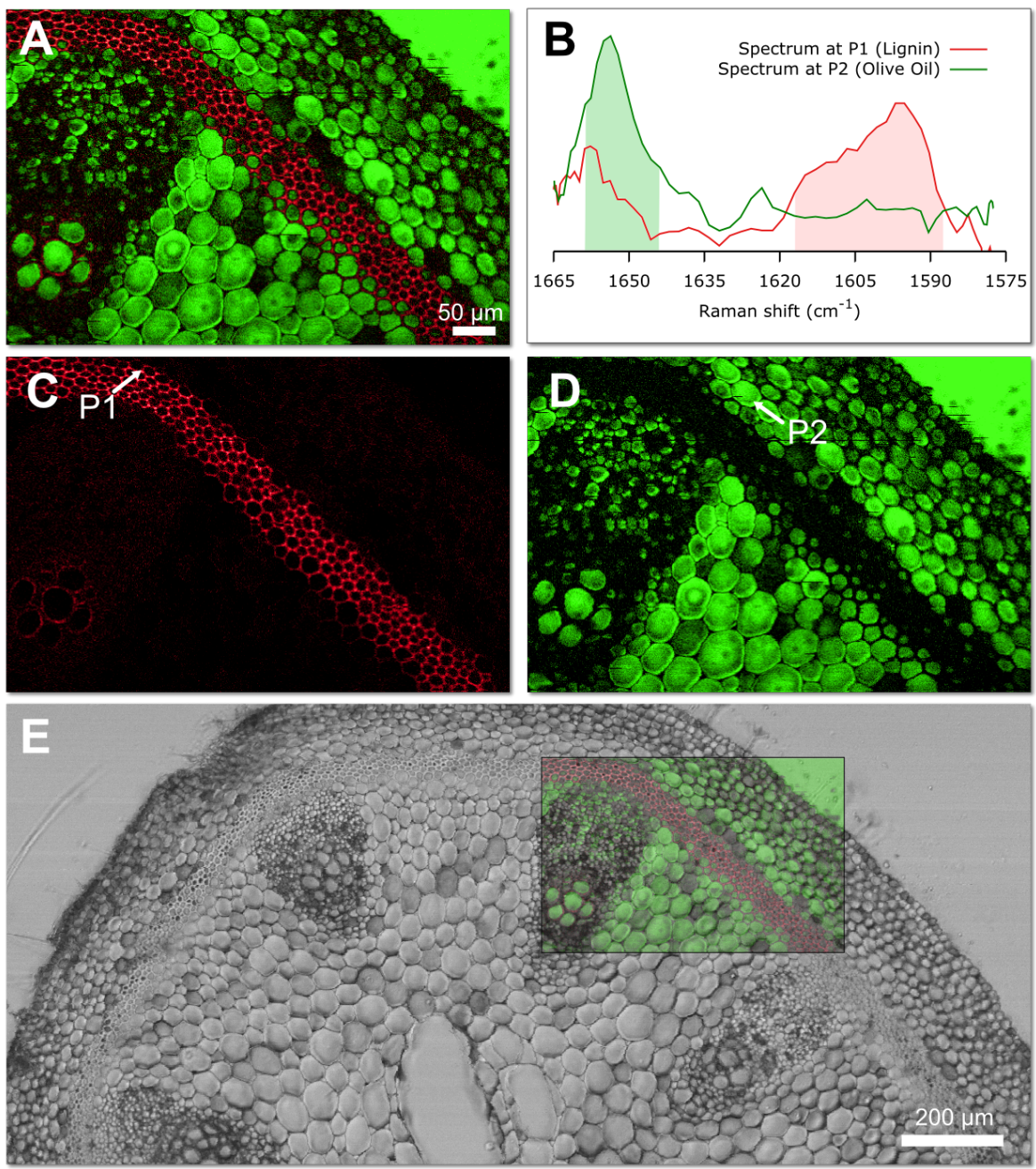

**Figure 4 Hyperspectral TICO-Raman images of a slice of geranium phaeum stem in olive oil.** (**A**) Image with molecular color code with lignin in red and olive oil in green. (**B**) Spectra at two pixels P1 and P2 and integration intervals used for color coding. (**C, D**) High signal-to-noise images of the two color channels. (**E**) Morphological overlay of the molecular contrast image from the probe laser with the high definition confocal image from the pump laser of the same setup.